\documentclass[conference]{IEEEtran}
\IEEEoverridecommandlockouts
\usepackage[english]{babel}
\usepackage[utf8]{inputenc}
\usepackage{cite}
\usepackage{amsmath,amssymb,amsfonts}
\usepackage{algorithmic}
\usepackage{graphicx}
\usepackage{textcomp}
\usepackage{xcolor}
\usepackage{caption}
\usepackage{subcaption}
\usepackage{tikz, tkz-euclide} 
\usepackage{mathrsfs}
\usetikzlibrary{shapes.geometric, shapes.symbols, calc, arrows, babel, intersections, patterns, patterns.meta, shapes, fit, positioning, arrows.meta, shapes.arrows}
\usetikzlibrary{positioning,fit,calc}
\usepackage[american,oldvoltagedirection]{circuitikz}
\usepackage{rotating}
\usepackage{listings}
\usepackage[hidelinks]{hyperref}
\usepackage[nameinlink,english,noabbrev]{cleveref}
\usepackage{float}
\usepackage{pgfplots}
\usepackage{url}
\usepackage{svg}
\usepackage{multirow}
\usepackage{multicol}
\usepackage{diagbox}
\usepgfplotslibrary{groupplots}
\usepackage{placeins}
\usepackage{comment}

\usepackage{orcidlink}

\pgfplotsset{compat=1.18}

\begin{document}
\renewcommand{\arraystretch}{1.2}
\setlength{\textfloatsep}{1\baselineskip plus 0.2\baselineskip minus 0.5\baselineskip}
%

\newcommand{\titlepaper}{Interpolation of Non-Linear Functions for LLMs using Partial Reconfiguration in FPGAs}

\title{\titlepaper}

\author{
  \IEEEauthorblockN{
    Roger de Jesús Morales Monge\orcidlink{0009-0009-2438-1745},
    Názareth Fabiola Jiménez Chacón\orcidlink{0009-0005-3465-1738},
    Jose Gabriel Villalobos Alvarado\orcidlink{0009-0003-8476-6619}, \\
    Luis~G.~Le\'on-Vega\orcidlink{0000-0002-3263-7853},
    Jorge~Castro-God\'{i}nez\orcidlink{0000-0003-4808-4904}
  }
  \vspace{3pt}
  \IEEEauthorblockA{
      Costa Rica Institute of Technology, Cartago, Costa Rica\\
  }
  \vspace{4pt}
  \IEEEauthorblockA{
    \{jgavillalobos, nazareth.jimenez, rj.m.m.2500\}@estudiantec.cr, \{l.leon, jocastro\}@tec.ac.cr
  }
}

\renewcommand{\abstractname}{Abstract}
\renewcommand\IEEEkeywordsname{Keywords}
\IEEEtitleabstractindextext{
\begin{abstract}
Non-linear functions such as exponential and sigmoid are essential in AI and LLM acceleration, although implementing them efficiently on FPGAs is still costly. This paper proposes a PWL interpolation framework based on partial reconfiguration to reduce hardware cost while preserving flexibility. The architecture separates the design into a static region for communication and control, and a reconfigurable region where different interpolation modules can be dynamically loaded. Uniform and non-uniform segmentation strategies are evaluated for exponential and sigmoid functions using FP16 and FP32 arithmetic. Results show that non-uniform segmentation can improve accuracy in high-curvature regions, while uniform segmentation offers lower hardware overhead. At the system level, the reconfigurable implementation achieved significant area savings, reaching up to 43\% less LUTs, 50\% less flip-flops, BRAMs and DSPs cells, compared against a static design containing both operators; all this with predictable reconfiguration latency. These results show that partial reconfiguration is a practical approach for exploring area-latency trade-offs in FPGA-based acceleration of non-linear functions for LLM workloads.
\end{abstract}

}

\maketitle

\IEEEdisplaynontitleabstractindextext

\IEEEpeerreviewmaketitle

\section{Introduction}

Large Language Models (LLMs) have advanced rapidly in recent years and gained widespread adoption due to their versatility in both research and everyday applications \cite{Li2024HWOrientedSoftmaxLayerNorm}. However, their high computational complexity and dependence on powerful hardware limit deployment in resource-constrained environments, motivating the development of specialized hardware accelerators.

Although most hardware acceleration efforts have traditionally focused on convolutions and matrix multiplication, modern Machine Learning (ML) architectures also rely on computationally demanding non-linear functions such as exponentiation, sigmoid, division, and trigonometric operations \cite{Andri2025Flex}. While linear algebra kernels are highly optimized in modern platforms, activation functions often remain less efficient because they are typically executed using more general-purpose resources.

Among these operations, softmax and layer normalization are especially relevant in LLMs, with prior work showing that they can account for a significant portion of GPU runtime \cite{peng2022length, stevens2021softermax}. In resource-constrained platforms such as FPGAs, these functions are particularly expensive to implement due the inherent computational complexity of non-linear operations such as exponential, square root and reciprocal. These operations often requires high amount of hardware resources to implement and achieve precise computations, which has increased interest in dedicated hardware accelerators. The challenge in these designs is to reduce hardware cost, latency, and power consumption while maintaining acceptable accuracy and inference performance.


\section{Related Work}

Approximate non-linear function implementations have been explored through several interpolation strategies. LUT-based methods combining Taylor-series approach achieve low approximation error and up to $20\%$ resource savings, though they often suffer from increased execution time~\cite{LeivaValverde2025ApproxSoftmax}. Piecewise Linear (PWL) designs such as Flex-SFU leverage non-uniform segmentation and SGF-based breakpoint selection to support multiple numeric formats, reporting up to $22.3\times$ accuracy improvement and $3.3\times$ inference speedup~\cite{Andri2025Flex}. Hybrid piece-wise polynomial Optimization (HyPPO) combines linear and quadratic segments under error constraints, achieving $18.14\%$ memory savings over PWL and $13\%$ over Piecewise Quadratic approach while maintaining accuracy~\cite{Vulchi2025HyPPO}. Other complementary optimization strategies, such as ILP -based coefficient rounding, reduce LUT requirements and enhance efficiency at higher precision~\cite{DeCaro2017ILPExactRounding}.

Beyond interpolation, Dynamic Partial Reconfiguration (DPR) has been explored to improve architectural flexibility. Prior work on reconfigurable CNN accelerators demonstrate that Partial Reconfiguration (PR) can effectively save in hardware costs and execution efficiency by enabling runtime adaptation~\cite{IrmakFPL2021DPRCNN}. Similarly, adaptive FPGA accelerators employ DPR to improve the utilization in ML workloads~\cite{ElBouazzaoui2024Adaptive}. While these works highlight the potential of DPR for balancing flexibility and performance, most studies remain focused on single optimization objectives, static hardware configuration or ASIC-oriented flows. This leaves FPGA-aware approximation strategies, particularly applied to LLM workloads, as an open research direction, where DPR and interpolation approach could better capture trade-offs among accuracy, resource usage, latency and flexibility.

Our contribution consists of a PWL accelerator for non-linear functions used in LLMs, implemented on a PR based architecture for FPGA platforms. We explored the trade-offs between area and latency in the deployment of hardware accelerators, while avoiding simultaneous instantiation of multiple interpolation engines.

\section{Proposed solution}

\begin{figure}[tbp]
    \centering
    \resizebox{0.99\linewidth}{!}{
    \begin{tikzpicture}[
    font=\Huge,
    >=Latex,
    block/.style={
        draw,
        thick,
        rounded corners=2mm,
        fill=blue!5,
        align=center,
        minimum width=3.2cm,
        minimum height=1.8cm
    },
    smallblock/.style={
        draw,
        thick,
        rounded corners=2mm,
        fill=blue!5,
        align=center,
        minimum width=3.0cm,
        minimum height=1.5cm
    },
    bigblock/.style={
        draw,
        thick,
        rounded corners=2mm,
        fill=blue!5,
        align=center,
        minimum width=4.0cm,
        minimum height=3.6cm
    },
    outerbox/.style={
        draw,
        thick,
        rounded corners=2mm,
        inner sep=10mm
    },
    outerboxdash/.style={
        draw=magenta,
        dashed,
        thick,
        rounded corners=2mm,
        inner sep=7mm
    },
    annot/.style={
        align=left,
        text width=4.2cm
    },
]

\definecolor{arrowfill}{RGB}{255,243,217}
\def\arrowinner{0.15}
\def\arrowouter{0.30}
\def\arrowoffset{0.50}

\node[block] (mpsoc) {Zynq\\Ultrascale+\\MPSoC};

\node[smallblock, below=3.0cm of mpsoc] (smartconnect) {AXI\\SmartConnect};

\node[smallblock, right=3.4cm of smartconnect] (shutdown) {AXI Shutdown\\Manager};

\node[smallblock, yshift=-2cm] at ($(smartconnect)!0.5!(shutdown)$) (gpio_shutdown) {AXI GPIO};


\node[bigblock, right=5.8cm of shutdown] (reconf) {Reconfigurable\\Region};

\node[annot, right=3cm of gpio_shutdown.east, yshift=-0.4cm] (note)
 {Enable/disable\\shutdown.};

\node[align=left, yshift=0.7cm, xshift=2.2cm] at (shutdown.east) (note2)
{data\_in/out};

\node[outerbox, fit=(mpsoc)(smartconnect)(shutdown)(gpio_shutdown)(note)(note2)] (staticbox) {};

\node[anchor=north east, font=\Huge\bfseries]
at ([xshift=-3mm,yshift=-2.5mm]staticbox.north east)
{Static Region};


\newcommand{\doublevertarrow}[2]{%
\fill[arrowfill, draw=black, thick, rounded corners=2pt]
(#1)
-- ($(#1)+(\arrowouter,-\arrowoffset)$)
-- ($(#1)+(\arrowinner,-\arrowoffset)$)
-- ($(#2)+(\arrowinner,\arrowoffset)$)
-- ($(#2)+(\arrowouter,\arrowoffset)$)
-- (#2)
-- ($(#2)+(-\arrowouter,\arrowoffset)$)
-- ($(#2)+(-\arrowinner,\arrowoffset)$)
-- ($(#1)+(-\arrowinner,-\arrowoffset)$)
-- ($(#1)+(-\arrowouter,-\arrowoffset)$)
-- cycle;
}

\newcommand{\doublehorarrow}[2]{%
\fill[arrowfill, draw=black, thick, rounded corners=2pt]
(#1)
-- ($(#1)+(\arrowoffset,\arrowouter)$)
-- ($(#1)+(\arrowoffset,\arrowinner)$)
-- ($(#2)+(-\arrowoffset,\arrowinner)$)
-- ($(#2)+(-\arrowoffset,\arrowouter)$)
-- (#2)
-- ($(#2)+(-\arrowoffset,-\arrowouter)$)
-- ($(#2)+(-\arrowoffset,-\arrowinner)$)
-- ($(#1)+(\arrowoffset,-\arrowinner)$)
-- ($(#1)+(\arrowoffset,-\arrowouter)$)
-- cycle;
}

\newcommand{\doublelsharrow}[2]{%
\fill[arrowfill, draw=black, thick, rounded corners=2pt]
(#1)
-- ($(#1)+(\arrowouter,-\arrowoffset)$)
-- ($(#1)+(\arrowinner,-\arrowoffset)$)
-- ($(#1|-#2)+(\arrowinner,\arrowinner)$)
-- ($(#2)+(-\arrowoffset,\arrowinner)$)
-- ($(#2)+(-\arrowoffset,\arrowouter)$)
-- (#2)
-- ($(#2)+(-\arrowoffset,-\arrowouter)$)
-- ($(#2)+(-\arrowoffset,-\arrowinner)$)
-- ($(#1|-#2)+(-\arrowinner,-\arrowinner)$)
-- ($(#1)+(-\arrowinner,-\arrowoffset)$)
-- ($(#1)+(-\arrowouter,-\arrowoffset)$)
-- cycle;
}

\newcommand{\doubleilsharrow}[2]{%
\fill[arrowfill, draw=black, thick, rounded corners=2pt]
(#1)
-- ($(#1)+(\arrowouter,-\arrowoffset)$)
-- ($(#1)+(\arrowinner,-\arrowoffset)$)
-- ($(#1|-#2)+(\arrowinner,-\arrowinner)$)
-- ($(#2)+(\arrowoffset,-\arrowinner)$)
-- ($(#2)+(\arrowoffset,-\arrowouter)$)
-- (#2)
-- ($(#2)+(\arrowoffset,\arrowouter)$)
-- ($(#2)+(\arrowoffset,\arrowinner)$)
-- ($(#1|-#2)+(-\arrowinner,\arrowinner)$)
-- ($(#1)+(-\arrowinner,-\arrowoffset)$)
-- ($(#1)+(-\arrowouter,-\arrowoffset)$)
-- cycle;
}

\doublevertarrow{mpsoc.south}{smartconnect.north}

\doublehorarrow{smartconnect.east}{shutdown.west}

\doublehorarrow{shutdown.east}{reconf.west}

\doubleilsharrow{shutdown.south}{gpio_shutdown.east}

\path ($(smartconnect.south)!0.5!(smartconnect.south-|smartconnect.east)$) coordinate (temp1);
\doublelsharrow{temp1}{gpio_shutdown.west}

\path ($(smartconnect.south)!0.5!(smartconnect.south-|smartconnect.west)$) coordinate (temp1);

\node[outerboxdash, fit=(mpsoc)] (ps) {};

\node[anchor=west, font=\Huge\bfseries]
at ([xshift=3mm]ps.east)
{Processing System (PS)};

\node[outerboxdash, fit=(smartconnect)(shutdown)(gpio_shutdown)(note)(note2)(reconf)] (pl) {};

\node[anchor=south, font=\Huge\bfseries]
at ([yshift=3mm]pl.north)
{Programmable Logic (PL)};

\end{tikzpicture}
    }
    \caption{Diagram of the proposed solution. The implementation consists of a reconfigurable region in which the interpolation unit is installed.}
    \label{fig:fpga_dfx_diagram}
\end{figure}
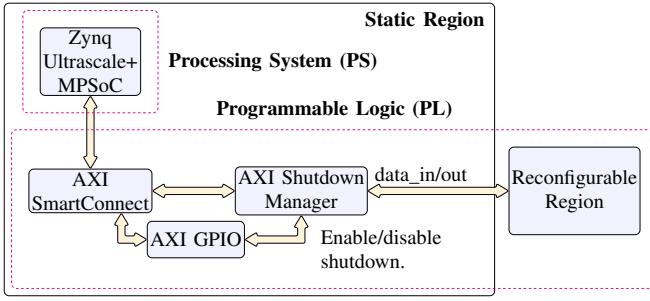

The proposed solution consists PR based architecture for non-linear functions interpolation engines. The design will be composed of a static region containing the shared data and control infrastructure, together with a reconfigurable region capable of loading different interpolation modules \cite{VipinFahmy2018PRSurvey}. \Cref{fig:fpga_dfx_diagram} shows the implemented architecture, the static region includes the Zynq UltraScale+ MPSoC (xck26-sfvc784-2LV-c), operating at a target frequency of 100 MHz, together with the AXI SmartConnect, AXI Shutdown Manager, and AXI GPIO as the backbone for communication and control. The reconfigurable region loads different interpolation engines depending on the non-linear function under evaluation.

Our proposal addresses area and flexibility challenges by employing a time-multiplexed accelerator architecture, which reduces required hardware without restricting the approximation to a single method. However, we recognize that PR introduces a non-negligible latency overhead due to the time required to deploy the partial bitstreams during reconfiguration, which makes it more suitable for coarse-grain changes rather than switching operations at token or layer level. Its viability lies on area and flexibility critical scenarios, when reconfiguration cost could be amortized over long execution interval \cite{LeivaValverde2025ApproxSoftmax, PrasadISVLSI2025PACE, Li2024HWOrientedSoftmaxLayerNorm}.

The scope of the solution is limited to the evaluation of FPGA-based PWL kernels for non-linear operations often used in  LLM acceleration. It focuses on reduced-precision arithmetic and interchangeable approximation engines, rather than the full LLM pipeline. The source code of the proposed solution is available at the 
\href{https://github.com/RMorales25/Non-LinearInterpolators_PartialReconfiguration}{GitHub repository}.



\section{Implementation}

\subsection{Interpolation Coefficients Generation}

The approximated function depends on the distribution and values of the interpolation coefficients, as they determine the terms of the linear equations that approximate the target function across a segmented input range. All functions were evaluated on the range of $[-8, 8]$, as representative range for the typical domain of the target functions within LLM models, including the high curvature regions and asymptotic tails of the sigmoid.

For a pair of breakpoints $x_1$ and $x_2$, the linear equation $f(x) \approx m (x+b/m)$ interpolates function $f(x)$ for the range $[x_1, x_2]$, where the slope is $m$ and the intercept value is $b$. A slope factorized form was used to avoid coefficient overflow on fast growing functions.




For uniform PWL strategy an equally spaced distribution of breakpoints is assigned across the input range, whereas for a non-uniform segmentation the breakpoints are allocated with more density on high curvature regions to allow smoother fit. A curvature-based approach is used to allocate non-uniform segments, proportionally to the second derivative~    \cite{Berjon_2016}.

\subsection{PWL Kernel Implementation}

\begin{figure}[tbp]
    \centering    
    \resizebox{0.99\linewidth}{!}{
        \begin{tikzpicture}[
    font=\huge,
    >=Latex,
    block/.style={
        draw,
        thick,
        rounded corners=2mm,
        fill=blue!5,
        align=center,
        minimum width=3.2cm,
        minimum height=1.8cm
    },
    smallblock/.style={
        draw,
        thick,
        rounded corners=2mm,
        fill=blue!5,
        align=center,
        minimum width=3.0cm,
        minimum height=1.5cm
    },
    outerboxdash/.style={
        draw=magenta,
        dashed,
        thick,
        rounded corners=2mm,
        inner sep=7mm
    },
    outerbox/.style={
    draw,
    thick,
    rounded corners=2mm,
    inner sep=10mm
    },
]

\definecolor{arrowfill}{RGB}{255,243,217}
\def\arrowinner{0.15}
\def\arrowouter{0.30}
\def\arrowoffset{0.50}

\node[block] (load) {Load input};
\node[block, below=2.24cm of load] (store) {Store output};

\node[outerboxdash, fit=(load)(store)] (staticbox) {};
\node[anchor=north east, font=\huge\bfseries]
at ([xshift=-3mm,yshift=-0mm]staticbox.north east)
{Static};

\node[smallblock, right=4.5cm of load] (unpack) {Unpacking};
\node[smallblock, right=3.2cm of unpack] (search) {Search\\Index};

\node[smallblock, below=2.5cm of unpack] (pack) {Packing};
\node[smallblock, right=3.2cm of pack] (interp) {Interpolate};

\node[outerbox, fit=(unpack)(search)(pack)(interp)] (compute) {};
\node[anchor=north east, font=\huge\bfseries]
at ([xshift=-3mm,yshift=-3mm]compute.north east)
{Compute};

\node[outerboxdash, fit=(compute)] (dynbox) {};
\node[anchor=south, font=\huge\bfseries]
at ([yshift=3mm]dynbox.north)
{Dynamic/Reconfigurable};


\newcommand{\doublevertarrow}[2]{%
\fill[arrowfill, draw=black, thick, rounded corners=2pt]
(#1)
-- ($(#1)+(\arrowouter,-\arrowoffset)$)
-- ($(#1)+(\arrowinner,-\arrowoffset)$)
-- ($(#2)+(\arrowinner,\arrowoffset)$)
-- ($(#2)+(\arrowouter,\arrowoffset)$)
-- (#2)
-- ($(#2)+(-\arrowouter,\arrowoffset)$)
-- ($(#2)+(-\arrowinner,\arrowoffset)$)
-- ($(#1)+(-\arrowinner,-\arrowoffset)$)
-- ($(#1)+(-\arrowouter,-\arrowoffset)$)
-- cycle;
}

\newcommand{\doublehorarrow}[2]{%
\fill[arrowfill, draw=black, thick, rounded corners=2pt]
(#1)
-- ($(#1)+(\arrowoffset,\arrowouter)$)
-- ($(#1)+(\arrowoffset,\arrowinner)$)
-- ($(#2)+(-\arrowoffset,\arrowinner)$)
-- ($(#2)+(-\arrowoffset,\arrowouter)$)
-- (#2)
-- ($(#2)+(-\arrowoffset,-\arrowouter)$)
-- ($(#2)+(-\arrowoffset,-\arrowinner)$)
-- ($(#1)+(\arrowoffset,-\arrowinner)$)
-- ($(#1)+(\arrowoffset,-\arrowouter)$)
-- cycle;
}

\newcommand{\doublelsharrow}[2]{%
\fill[arrowfill, draw=black, thick, rounded corners=2pt]
(#1)
-- ($(#1)+(\arrowouter,-\arrowoffset)$)
-- ($(#1)+(\arrowinner,-\arrowoffset)$)
-- ($(#1|-#2)+(\arrowinner,\arrowinner)$)
-- ($(#2)+(-\arrowoffset,\arrowinner)$)
-- ($(#2)+(-\arrowoffset,\arrowouter)$)
-- (#2)
-- ($(#2)+(-\arrowoffset,-\arrowouter)$)
-- ($(#2)+(-\arrowoffset,-\arrowinner)$)
-- ($(#1|-#2)+(-\arrowinner,-\arrowinner)$)
-- ($(#1)+(-\arrowinner,-\arrowoffset)$)
-- ($(#1)+(-\arrowouter,-\arrowoffset)$)
-- cycle;
}

\newcommand{\doubleilsharrow}[2]{%
\fill[arrowfill, draw=black, thick, rounded corners=2pt]
(#1)
-- ($(#1)+(\arrowouter,-\arrowoffset)$)
-- ($(#1)+(\arrowinner,-\arrowoffset)$)
-- ($(#1|-#2)+(\arrowinner,-\arrowinner)$)
-- ($(#2)+(\arrowoffset,-\arrowinner)$)
-- ($(#2)+(\arrowoffset,-\arrowouter)$)
-- (#2)
-- ($(#2)+(\arrowoffset,\arrowouter)$)
-- ($(#2)+(\arrowoffset,\arrowinner)$)
-- ($(#1|-#2)+(-\arrowinner,\arrowinner)$)
-- ($(#1)+(-\arrowinner,-\arrowoffset)$)
-- ($(#1)+(-\arrowouter,-\arrowoffset)$)
-- cycle;
}

\doublehorarrow{load.east}{unpack.west}

\doublehorarrow{store.east}{pack.west}

\doublehorarrow{unpack.east}{search.west}

\doublehorarrow{pack.east}{interp.west}

\doublevertarrow{search.south}{interp.north}

\node at ($(load.east)!0.4!(unpack.west)+(0,0.5)$) {\huge AXI};

\end{tikzpicture}
    }
    \caption{Diagram of the hardware accelerator for PWL interpolation. The reconfigurable region contains the unpacking/packing for handling different numerical precisions and the interpolation logic.}
    \label{fig:hw_acc_diagram}
\end{figure}
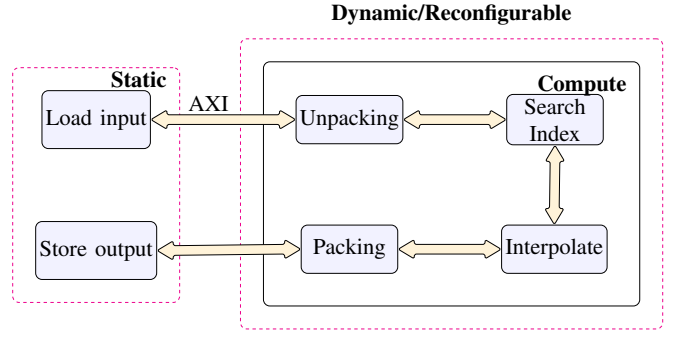

The PWL kernel was coded in high-level C++ language, which is then synthesized and compiled into a bitstream using Vitis HLS 2023.2 flow. \Cref{fig:hw_acc_diagram} shows the hardware accelerator, implemented as a standard load-compute-store structure. The load and store operations are executed in the static region, while the reconfigurable region holds compute stage, composed by the packing and unpacking operations, the index searching mechanism and the interpolation kernel.

All communication interfaces are implemented using AXI4 protocol for high bandwidth memory access. Data is streamed to the kernel in packets, then the compute stage performs interpolation based on the linear equations using the corresponding coefficients. 

For uniform PWL, the index is computed with arithmetic operations, while for non-uniform approach an index searching strategy is required. Since the classic comparator chain has poor scalability and suffers of high latency, a binary search approach is used. Then once the index is computed, the corresponding coefficients are loaded from memory and interpolation is done at the cost of one multiplication and one addition. All coefficients are stored in LUT-RAM memories, as a suitable alternative to BRAMs for small memory structures.

\subsection{Partial Reconfigurability and FPGA deployment}

The proposed system was implemented on the Kria KV60 platform, using a DFX architecture to preserve the static region while the interpolator are mapped to the reconfigurable partition. This enabled time-multiplexed execution of interpolation for each target function engine, loaded through reconfiguration using partial bitstreams and the Xilinx PR flow. 

PYNQ was employed as the software runtime for the accelerator deployment and execution on the platform embedded with Linux. Standard PYNQ flow was used to program the full bitstream for the baseline static configuration, while for DFX configuration the flow was further extended to support PR. The reconfiguration sequence was coordinated via MMIO registers to allow software controlled shutdowns, reset handshakes, hardware metadata export and control of the dynamic region. This demonstrates PYNQ can be used as a lightweight runtime environment for software-drive DFX management on FPGA platforms.

\section{Results and Analysis}

\begin{table}[t]
\centering
\captionof{table}{Error results for target functions approximation with PWL}
\label{tab:error-results}
\huge
\resizebox{\linewidth}{!}{%
\begin{tabular}{l|llccc}
    \hline
    \textbf{Function} & \textbf{Precision} &
    \textbf{PWL} &
    \textbf{Max Abs} &
    \textbf{Mean Abs.} &
    \textbf{Mean Rel. (\%)} \\
    \hline
    \hline
    \multirow{4}{*}{Exponential} & \multirow{2}{*}{FP32}
    & Uniform     & 1.41284 & 0.0607271 & 0.0337984 \\
    & & Non-Uniform & 0.172142 & 0.0188023 & 2.22582 \\
    \cline{2-6}
    & \multirow{2}{*}{FP16}
    & Uniform     & 2 & 0.036578 & 0.0164993 \\
    & & Non-Uniform & 6 & 0.177017 & 1.36598 \\
    \hline
    \hline
    \multirow{4}{*}{Sigmoid} & \multirow{2}{*}{FP32}
    & Uniform     & 0.000783145 & 4.83E-05 & 0.0178621 \\
    & & Non-Uniform & 1.28E-05 & 5.87E-06 & 0.064295 \\
    \cline{2-6}
    & \multirow{2}{*}{FP16}
    & Uniform     & 0.00146484 & 0.000159925 & 0.0263574 \\
    & & Non-Uniform & 0.000976562 & 0.000130281 & 0.0217006 \\
    \hline
\end{tabular}%
}
\end{table}

To evaluate the proposed solution, two experimental architectures are considered. Baseline implementation instantiates separate interpolation engines, both coexisting in the FPGA fabric simultaneously, while the PR implementation instantiates only one interpolator at any time, allowing to switch target function through the reconfiguration process. 

To ensure interpolator produce valid approximations of the target function, the accuracy is measured through absolute and relative error metrics. Then FPGA area utilization (LUTs, DSPs, FF, BRAMs), power consumption, compute latency are measured directly during execution on the Kria platform, to evaluate the trade-off between resources and performance. Additionally, reconfiguration latency is measured for DFX configuration, as the time required to switch between kernels. 

Power measurements were sample periodically from sensors exposed to the embedded Linux subsystem. For all experiments the baseline power consumption difference is measured between the idle state and the active window, during computation and reconfiguration process. This way, power results emphasize the incremental energy cost of the DFX approach.

For latency measurements, dedicated hardware counters are implemented in the programmable fabric of the FPGA and exposed through AXI GPIO. This allows direct hardware measures, instead of relying on host-side timestamps, which provides consistent view of the system and accurate measures under realistic board conditions.

\Cref{tab:error-results} shows error results for uniform and non-uniform interpolation for exponential and sigmoid functions respectively. Error was evaluated during cosimulation with the synthesized design for both FP16 and FP32 implementations. Error metrics were computed against output values of the reference function quantized to the target numeric precision.


FP32 generally reduces absolute error, while relative error depends on segmentation strategy. For the exponential, non-uniform segmentation improves absolute error by allocating more breakpoints in high-value regions, but increases relative error due to sensitivity in low magnitude outputs. For sigmoid, non-uniform segmentation concentrates breakpoints in the high curvature region near zero, improving absolute error, with a limited impact on the tails. 

In FP16, limited mantissa precision leads to coarser quantization for higher output values, increasing absolute error, while relative error is dominated by finer quantization on low value outputs. Thus, segmentation strategies that prioritize high-value regions tend to worsen relative error on high-curved functions such as the exponential.

Overall, non-uniform segmentation distributes error across the domain. The effectiveness of non-uniform segmentation depends on target function and error optimization objective. Sigmoid benefits from this trade-off, while not for exponential in a wide domain. Nevertheless, non-uniform segmentation offers flexibility, allowing balance between accuracy and hardware cost as breakpoints could be potentially reduced for a given error.

\Cref{tab:compact_power,tab:compact_area} report post-implementation power, area and latency results, respectively, for the selected static and reconfigurable configurations. This corresponds to validation on the target platform, focusing on the differences between baseline and DFX configurations.

\begin{table}[t]
\centering
\caption{Comparison of power delta for FP32 and FP16 implementations.}
\label{tab:compact_power}
\setlength{\tabcolsep}{3pt}
\begin{tabular}{l|llcc}
\hline
\textbf{Precision} &
\textbf{Region} &
\textbf{DUT} &
\textbf{Reconf. Delta [W]} &
\textbf{Active Delta [W]} \\
\hline
\hline
\multirow{4}{*}{FP32} & \multirow{2}{*}{Static}
          & exp     & --       & 0.212298 \\
&         & sigmoid & --       & 0.214478 \\
& \multirow{2}{*}{Reconf.}
          & exp     & 0.408671 & 0.227367 \\
&         & sigmoid & 0.399669 & 0.225965 \\
\hline
\hline
\multirow{4}{*}{FP16} & \multirow{2}{*}{Static}
          & exp     & --       & 0.211779 \\
&         & sigmoid & --       & 0.210791 \\
& \multirow{2}{*}{Reconf.}
          & exp     & 0.415674 & 0.224544 \\
&         & sigmoid & 0.423660 & 0.224157 \\
\hline
\end{tabular}%
\end{table}

Relative to the static baseline, the active power for reconfiguration increases by $7.10\%$ for FP32 $\exp$, $5.36\%$ for FP32 sigmoid, $6.03\%$ for FP16 $\exp$, and $6.34\%$ for FP16 sigmoid. Despite this increase, all reconfigurable cases consume less than $0.02W$ over baseline. The partial reconfiguration latency remains in a narrow range between $2.71\times10^{6}$ and $2.80\times10^{6}$ cycles, indicating consistent runtime behavior. In terms of kernel latency, the FP32 reconfigurable implementations are consistently lower than their static counterparts, showing reductions close to $9\%\text{--}10\%$ across the 16k, 32k, and 64k input sizes, whereas the FP16 reconfigurable implementations exhibit only a small increase, below $1\%$ in all evaluated cases. Across all configurations, the design maintains stable efficiency profile under reconfiguration, with a moderate power overhead and predictable latency reconfiguration cost.

\begin{table}[t]
\centering
\caption{Comparison of  reconfiguration latency and application latency for FP32 and FP16 implementations at input sizes of 16k, 32k, and 64k samples.}
\label{tab:compact_latency}
\setlength{\tabcolsep}{3pt}
\large
\resizebox{\columnwidth}{!}{%
\begin{tabular}{l|llcccc}
\hline
\multirow{2}{*}{\textbf{Precision}} &
\multirow{2}{*}{\textbf{Region}} &
\multirow{2}{*}{\textbf{DUT}} &
\multirow{2}{*}{\textbf{Reconf. Latency}} &
\multicolumn{3}{c}{\textbf{Kernel Latency [cycles]}} \\
& & & \textbf{[cycles]} & \textbf{1}\textbf{6k} & \textbf{32k} & \textbf{64k} \\
\hline
\hline
\multirow{4}{*}{FP32} & \multirow{2}{*}{Static}
          & exp     & --         & 21345.20 & 41824.35 & 82784.00 \\
&         & sigmoid & --         & 21338.40 & 41815.60 & 82781.95 \\
& \multirow{2}{*}{Reconf.}
          & exp     & 2763412.70 & 19444.05 & 37814.60 & 74671.05 \\
&         & sigmoid & 2707559.05 & 19398.10 & 37807.55 & 74670.80 \\
\hline
\hline
\multirow{4}{*}{FP16} & \multirow{2}{*}{Static}
          & exp     & --         & 17269.45 & 33647.45 & 66369.35 \\
&         & sigmoid & --         & 17266.40 & 33621.70 & 66362.65 \\
& \multirow{2}{*}{Reconf.}
          & exp     & 2796313.30 & 17422.55 & 33774.00 & 66484.65 \\
&         & sigmoid & 2801157.45 & 17424.20 & 33846.55 & 66485.70 \\
\hline
\end{tabular}%
}
\end{table}

The DFX configuration provides a substantial reduction in area utilization, compared to the static architecture that instantiates both DUTs simultaneously. For FP32 precision, the reconfigurable implementation occupies 8488 LUTs, 4914 FFs, 1 BRAM, and 5 DSPs, whereas the combined static design uses 16438 LUTs, 9765 FFs, 2 BRAMs, and 10 DSPs, corresponding to reductions of $48.4\%$ in LUTs, $49.7\%$ in FFs, $50.0\%$ in BRAMs, and $50.0\%$ in DSPs. Similarly, for FP16 the DFX design requires 4532 LUTs, 4001 FFs, 1 BRAM, and 4 DSPs, while static implementation requires 8440 LUTs, 7978 FFs, 2 BRAMs, and 8 DSPs, corresponding savings of $46.3\%$ in LUTs, $49.8\%$ in FFs, $50.0\%$ in BRAMs, and $50.0\%$ in DSPs. Therefore, under a system-level comparison in which both interpolators are required, the DFX approach demonstrates better area efficiency, by the time-multiplexing of each kernel implementations with a shared reconfigurable region.


\section{Conclusions and Future Work}

This work introduced a partial reconfiguration based FPGA architecture for efficient non-linear function PWL interpolators in LLM accelerators. By enabling time-multiplexed interpolation engines, the proposed design achieved reduced hardware utilization, up to $43.1\%$ LUTs, $49.8\%$ and $50\%$ BRAMs and DSPs compared to static implementation, while preserving acceptable approximation accuracy. Although partial reconfiguration offers significant savings in resources usage, it introduces a non-neglible latency overhead, which could be amortized in coarse-grain switching scenarios, where reconfiguration process can be masked by other LLM operations.

Future work includes extending the partial reconfiguration approach to additional approximation methods and target functions, improving interpolation accuracy with numeric precision aware non-uniform segmentation, and scaling it to full LLM pipelines. Future research will also explore lower precision formats and reconfiguration latency optimization.

\begin{table}[t]
\centering
\caption{Comparison of area utilization post placement and route for the FP32 and FP16 implementations.}
\label{tab:compact_area}
\small
\setlength{\tabcolsep}{3pt}
\begin{tabular}{l|llcccc}
\hline
\textbf{Precision} & \textbf{Region} & \textbf{DUT} & \textbf{LUT} & \textbf{FF} & \textbf{BRAM} & \textbf{DSP} \\
\hline
\hline
\multirow{3}{*}{FP32} & \multirow{2}{*}{Static} & exp & 8116 & 4895 & 1 & 5 \\
& & sigmoid & 8322 & 4870 & 1 & 5 \\
& Reconf. & exp/sigmoid & 8488 & 4914 & 1 & 5 \\
\hline
\hline
\multirow{3}{*}{FP16} & \multirow{2}{*}{Static} & exp & 4155 & 3980 & 1 & 4 \\
& & sigmoid & 4285 & 3998 & 1 & 4 \\
& Reconf. & exp/sigmoid & 4532 & 4001 & 1 & 4 \\
\hline
\end{tabular}
\end{table}

\section*{Acknowledgments}

This work was supported by the Costa Rica Institute of Technology under the research project 1360058 (Generación Automática de Hardware para Aplicaciones de Aprendizaje Automático basadas en FPGA). The results of this work were partially supported by AMD through the Heterogeneous Accelerated Compute Clusters (HACC) program.

Additionally, AI-based tools were used exclusively for grammar correction and translation during the preparation of this manuscript.

\bibliographystyle{IEEEtran}
\bibliography{Paper}

\end{document}